\documentclass[a4paper]{spie}  

\usepackage{aas_macros}
\usepackage{amsmath,amsfonts,amssymb}
\usepackage{graphicx}
\usepackage[colorlinks=true, allcolors=blue]{hyperref}
\usepackage{svg}
\usepackage{lipsum}

\title{Application of continuous corrections in wide‑field imaging obtained with the IDG‑CAL calibration method}

\author{Sebastiaan van der Tol}
\affil{Netherlands Institute for Radio Astronomy (ASTRON), Postbus 2, 7990 AA Dwingeloo, The Netherlands}

\authorinfo{Author e-mail: tol@astron.nl}

\begin{document} 
\maketitle

\begin{abstract}
   Image Domain Gridding (IDG) is an efficient method to evaluate A-projection, a method for
   correcting for direction dependent effects (DDEs) in radio astronomical imaging.
   The corrections i.e. the A-terms need to be estimated from the observed data 
   through calibration. IDG-CAL is a calibration algorithm that employs 
   the IDG algorithm in an iterative fashion to estimate A-terms directly.
   Initial results for low order A-terms are promising, however for
   deeper images higher order A-terms are needed.
   Here two improvements to IDG-CAL are proposed to 1) reduce the computational cost
   of A-terms described by more parameters and 2) optimize the set of basis
   functions describing the A-terms based on a stochastic model of the gain variations
   and the model image.
\end{abstract}


\section{INTRODUCTION}
\label{sec:intro}  

In this paper we report on the ongoing development of IDG-CAL, a calibration algorithm to obtain
corrections for direction dependent effects in wide field imaging of radio interferometric data.
A detailed description of the algorithm can be found in Ref. \citenum{vandertol2026}.
Here we give a more general intuitive description of the ideas behind the algorithm.
We report on the current status, including the issues encountered when applying the algorithm to 
a full scale dataset. 
We present the mathematical derivation of two possible improvements for greater speed and better quality.

\begin{figure} [ht]
   \begin{center}
      \begin{tabular}{cc} 
         \includegraphics[height=5cm]{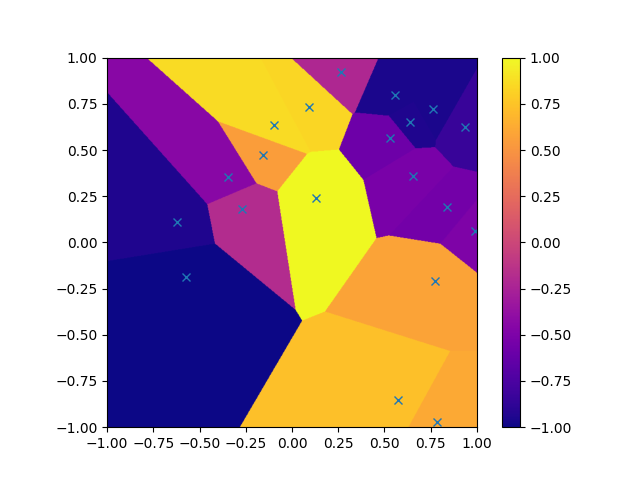} & \includegraphics[height=5cm]{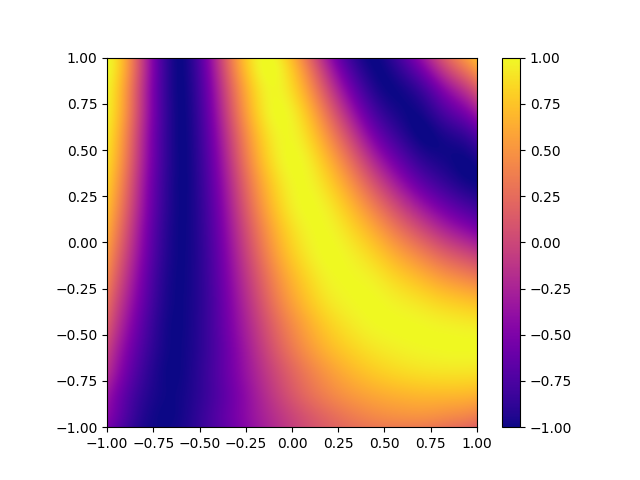}
      \end{tabular}
   \end{center}
   \caption
   { \label{fig:dde_methods} 
      Arbitrary example of a facet based versus a screen based correction.
      Facet based methods apply piece-wise constant correction. 
      A-projection applies an interpolated correction.
      The corrections are complex phasors, of which the real part is plotted.
   }
\end{figure}

The methods for correcting dependent effects can be divided into two categories.
The first category are facet based approaches\cite{weeren2016}, where the image is divided into smaller sub-images or facets.
The gains for each facet can than be treated as if the are direction independent.
The other approach is to apply the direction dependent effect as part of the the gridding kernel \cite{bhatnagar2008}.
Due to properties of the Fourier transform the effectively applied correction is an interpolation of 
the direction dependent gain sampled on a regular grid. Figure \ref{fig:dde_methods} illustrates the difference between the two methods.
The smooth corrections by gridding kernel based methods are physically more plausible than
the piece-wise constant corrections of facet based methods.
However, adoption of these methods has been limited to hampered by the difficulty of obtaining 
suitable corrections.

\section{THE IDG-CAL ALGORITHM}

\begin{figure} [ht]
   \begin{center}
   \begin{tabular}{c} 
   \includegraphics[height=0.7\textheight]{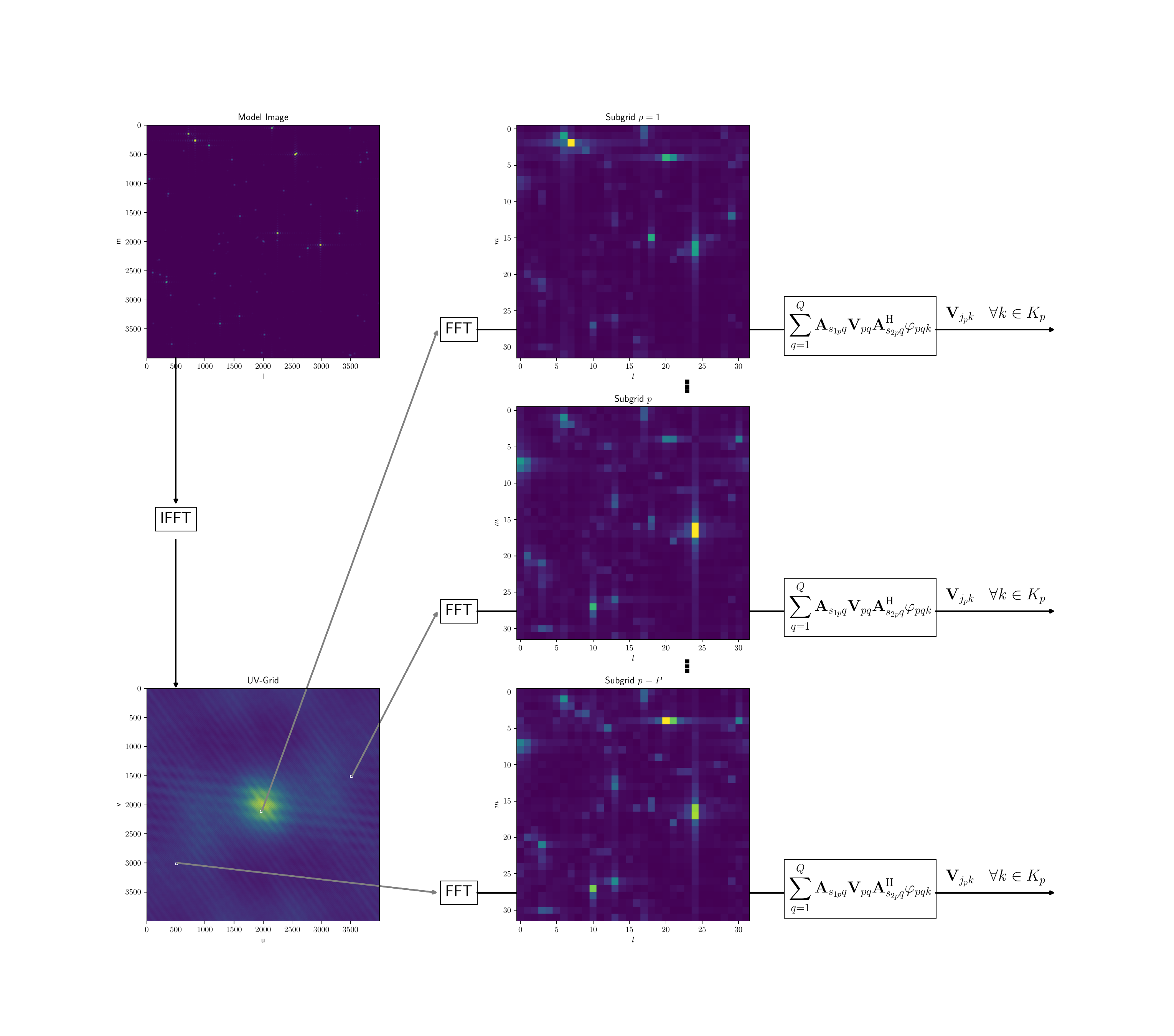}
   \end{tabular}
   \end{center}
   \caption[example] 
   { \label{fig:subgrids} Overview of IDG predict. The procedure consists of the following steps 
   1) the model image is inverse Fourier transformed to a uv-grid, 2) from the uv-grid sub-grids are 
   extracted, 3) the sub-grids are Fourier transformed back to the image domain, 4) the 
   predicted visibilities are computed as a weighted sum of the sub-grid pixels.
   In practice there are thousands of sub-grids. Here only three are shown.
   Each subgrid is a low resolution image of the sky as seen by that specific baseline.
   }
\end{figure}

Figure \ref{fig:subgrids} shows an overview of the IDG algorithm to predict visibilities.
The difference between the predicted visibilities and the observed visibilities are the residual
visibilities.

The idea behind IDG-CAL is to use the IDG algorithm for calibration.
This is done by parameterizing the A-terms and computing derivatives of those A-terms
with respect to the parameters. These can then be substituted in the IDG degridding
algorithm to predict visibilities and derivative visibilities.
From these the derivative of the cost function, the mean squared residual visibilities,
 can be computed. A generic solver can then estimate A-terms by minimizing this cost function.

The parameterization of the A-terms is done as follows.
The A-terms consist of two components: 1) a fast varying scalar phase and 2) a slow varying
complex diagonal matrix. In total the expression for the A-term is
\begin{equation}
   \mathbf{J}(\mathbf{p})_{jkq} = e^{\varphi(\mathbf{p}_{\mathrm{ph},k})\jmath}
   \left[
   \begin{array}{cc}
      a_1(\mathbf{p}_{\mathrm{a1}}) e^{\varphi(\mathbf{p}_{\mathrm{ph1}})\jmath} & \\
          & a_2(\mathbf{p}_{\mathrm{a2}}) e^{\varphi(\mathbf{p}_{\mathrm{ph2}})\jmath}
   \end{array}
   \right]
\end{equation}
Each component is written as a sum of basis functions $b(l,m)$:
\begin{equation}
   \varphi(\mathbf{p}_{\mathrm{ph},k}) = \sum_{r=1}^{R} p_{\mathrm{ph},kr} b_r(l,m)
\end{equation}
likewise for $a_1(\mathbf{p}_{\mathrm{a1}})$, $a_2(\mathbf{p}_{\mathrm{a1}})$, 
$\varphi(\mathbf{p}_{\mathrm{ph1}})$ and $\varphi(\mathbf{p}_{\mathrm{ph2}})$.

\section{CURRENT STATUS OF IDG-CAL}

In Ref. \citenum{vandertol2026} IDG-CAL was applied to a dataset of 16MHz bandwidth and a subset of 20\% of an 8 hour observation.
This resulted in a lower background noise compared to a run using 34 facets. 
The next step is to use the full bandwidth and the full time range.

Processing a full dataset (48Mhz, 8hours) is feasible, but at the border of what is practical.
Processing a block of 3MHz in frequency and 8 hours in time with second order polynomials (6 degrees of freedom)
on a high end machine (AMD EPYC 9555P 64-Core Processor) takes about 3.7 hours. Processing the full bandwidth then takes nearly 60 hours.

Compared to a faced based calibration and imaging run on the same dataset with 50 facets,
the resulting background noise was slightly higher and most sources have higher artifacts.

A likely explanation for why the initial promising results on the smaller dataset does
not translate to better results on the full dataset is the limited number of free parameters.
The IDG-CAL screen was described by only 6 coefficients, compared to 50 facets.
Because of the inherent smoothness of the screens compared to piece-wise constant facet based corrections,
it can be expected that fewer coefficients are needed then the number of facets in a comparable run.
In this case the 6 coefficients were probably not enough to track the gain fluctuations as accurate 
as is possible with 50 facets.

Scaling up the number of parameters with the current implementation is problematic for following reasons:
\begin{itemize}
   \item Compute time scales with the number of free parameters
   \item The polynomial base is not suitable, especially the higher order terms, which are very steep at the edges.
\end{itemize}

The next section describes two possible approaches to tackle these problems.

\section{PROPOSED IMPROVEMENTS}

\begin{figure} [t]
   \begin{center}
      \includesvg[width=\textwidth]{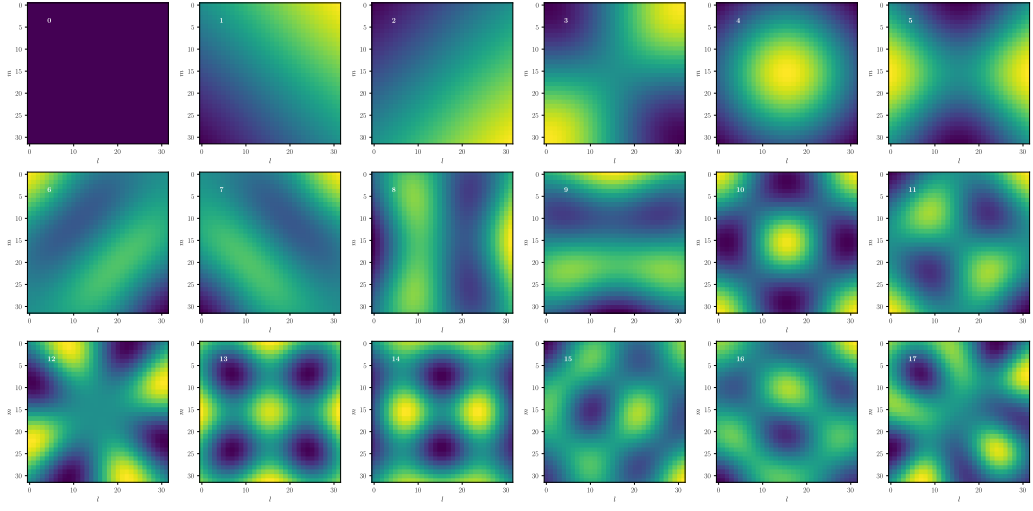}
   \end{center}
   \caption
   { \label{fig:kl-uniform} 
      2D Karhunen-Loève basis functions over $(l,m)$ for a uniform error.
      These basis function closely resemble Zernike polynomials.
   }
\end{figure} 
\begin{figure} [t]
   \begin{center}
      \begin{tabular}{c} 
         \includesvg[width=\textwidth]{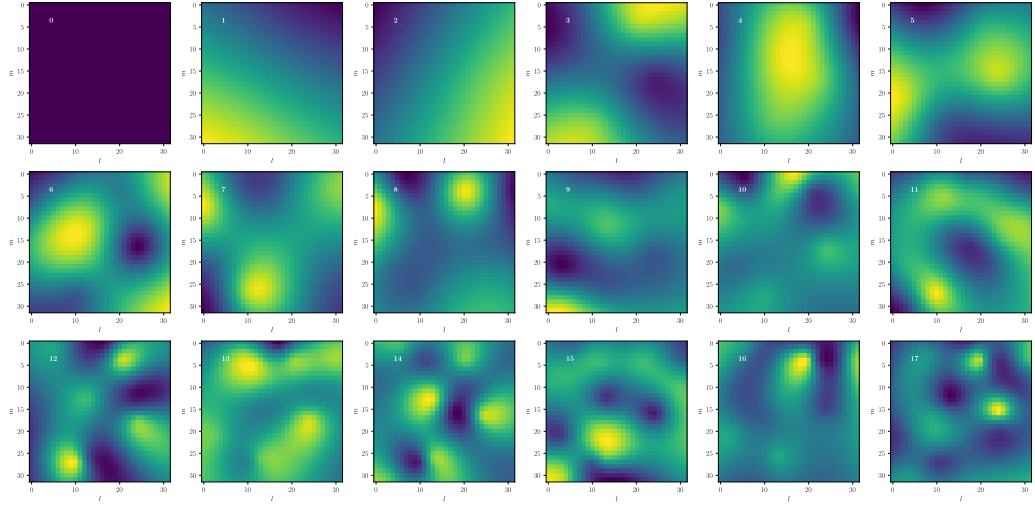}
      \end{tabular}
   \end{center}
   \caption[example] 
   { \label{fig:kl-weighted} 
      2D Karhunen-Loève basis functions over $(l,m)$, error weighted by the model image.
      The higher order functions are more detailed at the brighter sources
      compared to areas with little flux.
   }
\end{figure} 

To apply IDG-CAL to deep images, the screen needs to be described by more parameters.

To make this feasible two improvements are needed. The first is lowering the computational cost 
for problems with a large number of free parameters. That can be achieved by rearranging the equations
and evaluating them in a different order. The loop over the free parameters is thus kept outside
the inner loop.

The second improvement is using other base functions than polynomials, which are steep at the edged.
An optimal set of base functions is derived for a stochastic model of the gain variations.

\subsection{Reordering the computations}
Currently the runtime scales linearly with the number of free parameters.
It turns out that this does not need to be the case.
A suitable reordering of the computations can remove the loop over the free parameters
out of the critical loop.

The derivation starts with Eq. (7) from Ref. \citenum{vandertol2026}, where the derivative of the cost function is given as
a sum over all contributing sub-grids. The per sub-grid contribution is given by Eq. (14) from \citenum{vandertol2026}.
Merging those equations yields:
\begin{equation}
  \frac{\partial C(\mathbf{x})}{\partial x_{sr}}  =  -2 \operatorname{Re} \sum_{p \in P_s}\sum_{q=1}^Q \operatorname{vec}\left(\mathbf{V}_{pq}\right)^{\mathrm{H}} 
  \operatorname{vec}\left(
    \frac{\partial}{\partial x_{sr}}\mathbf{A}_{sq}^\mathrm{H}\mathbf{X}_{pq}\mathbf{A}_{s_{2p}q}
  \right) \\
\end{equation}
where $Q$ is the number of subgrid pixels, and $P_s$ is the set of sub-grids in which stations $s$
participates, $x_{sr}$  is the $r$th parameter of station $s$, $\mathbf{A}$ are the A-terms,
$\mathbf{V}$ are the model sub-grid pixels, $\mathbf{X}$ are the accumulated residual visibilities.
For more detail see Ref. \citenum{vandertol2026}.

Using the identity $\operatorname{vec}(\mathbf{ABC}) = \left(\mathbf{C}^\mathrm{T} \otimes \mathbf{A}\right) \operatorname{vec}\left(\mathbf{B}\right)$
this ran be rearranged into
\begin{eqnarray}
  \frac{\partial C(\mathbf{x})}{\partial x_{sr}}
  & = & -2 \operatorname{Re} \sum_{p \in P_s} \sum_{q=1}^Q \operatorname{vec}\left(\mathbf{V}_{pq}\right)^{\mathrm{H}}   \left(\left(\mathbf{X}_{pq}\mathbf{A}_{s_{2p}q}\right)^\mathrm{T}\otimes \mathbf{I}\right) \operatorname{vec}\left(\frac{\partial}{\partial x_{sr}}\mathbf{A}_{sq}^\mathrm{H}\right) \\
  & = & -2 \operatorname{Re} \sum_{p \in P_s}\sum_{q=1}^Q \left(   \left(\left(\mathbf{X}_{pq}\mathbf{A}_{s_{2p}q}\right)^\mathrm{*}\otimes \mathbf{I}\right) \operatorname{vec}\left(\mathbf{V}_{pq}\right)\right)^{\mathrm{H}} \operatorname{vec}\left(\frac{\partial}{\partial x_{sr}}\mathbf{A}_{sq}^\mathrm{H}\right)\\
  & = & -2 \operatorname{Re} \sum_{p \in P_s}\sum_{q=1}^Q \operatorname{vec}\left(\mathbf{V}_{pq} \mathbf{A}_{s_{2p}q}^\mathrm{H} \mathbf{X}_{pq}^\mathrm{H}  \right)^{\mathrm{H}} \operatorname{vec}\left(\frac{\partial}{\partial x_{sr}}\mathbf{A}_{sq}^\mathrm{H}\right).
\end{eqnarray}

Now define $\mathbf{Y}_{sq}$, the derivative of the cost function with respect to pixel $q$ as a sum over sub-grids:
\begin{equation}
  \mathbf{Y}_{sq} = \sum_{p \in P_s} \mathbf{V}_{pq} \mathbf{A}_{s_{2p}q}^\mathrm{H} \mathbf{X}_{pq}^\mathrm{H}.
\end{equation}
The derivative of the cost function with respect to parameter $x_{sr}$ can than be written as a sum over the pixels of the 
accumulated sub-grids
\begin{equation}
   \frac{\partial C(\mathbf{x})}{\partial x_{sr}} = -2 \operatorname{Re} \sum_{q=1}^{Q} \operatorname{vec}\left(\mathbf{Y}_{sq} \right)^{\mathrm{H}} \operatorname{vec}\left(\frac{\partial}{\partial x_{sr}}\mathbf{A}_{sq}^\mathrm{H}\right).
\end{equation}

The difference with the current implementation is that sub-grids will be accumulated over
stations before entering the loop over free parameters.
Especially for the computation of the derivatives of the long time scale complex diagonal term this rearrangement is advantageous.
The complex diagonal consists of four terms, two amplitudes and two phases, and each of them is described by $R$ basis functions.
By accumulating sub-grids first before looping over these terms, the contribution of this loop to the total cost is greatly reduced,
even for large $R$.

\subsection{An optimal set of base functions}
The polynomial functions are easy to implement and the lower order terms
give a decent description of the screen.
However, a better set of basis functions can be derived when modeling the fluctuations as
a stochastic process and then derive the optimal set of basis functions for that process.
These functions are known in literature as the Karhunen-Loève functions \cite{Jorgensen2007}.

The correlation function for the stochastic process is derived from Kolmogorov turbulence,
which is applicable to atmospheric effects \cite{intema2009}. For instrument based effects this model
is not an accurate description of the physics, although the resulting basis functions
might still be good enough. Otherwise a better stochastic model of the instrumental
variations can readily be substituted into the derivation below.

When assuming Kolmogorov turbulence in the atmosphere as the source 
of phase variations the structure function is given by:
\begin{equation}
   \operatorname{E} \left[ \left(x(\mathbf{p}_1) - x(\mathbf{p}_2)\right)^2 \right] = 
   \left\| \mathbf{p}_1 - \mathbf{p}_2 \right\|^\beta
\end{equation}

From the equation above the cross correlation function can be derived, except
for an unknown constant offset:
\begin{equation}
   \operatorname{E} \left[ \left(x(\mathbf{p}_1) - x(\mathbf{p}_2)\right)^2 \right] = 
   \operatorname{E} \left[ x^2(\mathbf{p}_1)\right] + \operatorname{E} \left[ x^2(\mathbf{p}_2) \right] -2 \operatorname{E} \left[  x(\mathbf{p}_1) x(\mathbf{p}_2) \right]
\end{equation}
For a structure function with no outer scale the autocorrelation is infinite. Here we ignore the auto correlations, and project the constant term
out of the correlation matrix. Afterwards a constant valued zeroth order basis function is added to the set.

The correlation matrix is given by
\begin{equation}
   \mathbf{C} = \frac{-1}{2} \left[
      \begin{array}{cccc}
         0 & \left((l_1 - l_2)^2 (m_1 - m_2)^2\right)^{\beta/2} & \hdots & \left((l_1 - l_Q)^2 (m_1 - m_Q)^2\right)^{\beta/2} \\
         \left((l_2 - l_1)^2 (m_2 - m_1)^2\right)^{\beta/2}  & 0 & & \\
         \vdots & & \ddots & \\  
         \left((l_Q - l_1)^2 (m_Q - m_1)^2\right)^{\beta/2} & & & 0
      \end{array}
   \right]
\end{equation}

The projection matrix removing the constant term is given by
\begin{equation}
   \mathbf{P} = \mathbf{I} - \mathbf{1}\left(\mathbf{1}^\mathrm{T}\mathbf{1}\right)^{-1}\mathbf{1}^\mathrm{T}
\end{equation}

The eigenvalue decomposition of the projected correlation matrix is given by
\begin{equation}
   \mathbf{PCP}^\mathrm{T} = \mathbf{USU}^\mathrm{T}
\end{equation}
The columns of $\mathbf{U}$ form a set of basis functions.
The basis functions are shown in Fig. \ref{fig:kl-uniform} including the added zeroth order constant term.

The error that is minimized by using this basis, is the sum over all pixels of the squared error, whereby all pixels get the same weight.
However, errors in a pixel with more flux contribute more to the error in the predicted visibilities.
To accommodate that effect a weighting by flux needs to be applied to the correlation matrix.
The flux weighted correlation matrix is given by
\begin{equation}
   \mathbf{C}^{\prime} = \mathbf{WCW}^\mathrm{T}
\end{equation}
where $\mathbf{W}$ is a diagonal matrix with the root mean squared flux on the diagonal.
The mean can be taken over all sub-grids, or over all sub-grids for a specific station, if per station basis functions are required.
The columns of $\mathbf{U}^{\prime}$ then contain the weighted basis functions.
The get the unweighted basis functions the weighting need to be divided out.
The division does not preserve the orthonormality. The orthonormal set of unweighted basis functions can be constructed by a QR-decomposition
as follows
\begin{equation}
   \mathbf{W}^{-1}\mathbf{U}^{\prime} = \mathbf{Q}\mathbf{R}
\end{equation}
The columns of $\mathbf{Q}$ now contain the basis functions.
The result is show in Fig. \ref{fig:kl-weighted}. The functions are smooth but show a bit more detail at the location of the brighter sources,
reserving a bit more freedom to follow fluctuations at the brighter sources without distorting the overall smooth solution.

 A trade off is made between having an independent solution for the brightest sources, as is possible in the facet based approach,
 and on the other side enforcing smoothness as is done by using screens described by unweighted KL-functions.
 
\section{CONCLUSIONS}
The initial promising results of IDG-CAL on smaller datasets do not carry over to 
larger datasets. A likely cause is the limited number (6) of free parameters.
Increasing the number of free parameters requires a reduction of run-time 
for cases with a large number of parameters. This can be achieved by changing
to order of the loops over station and parameters.
Also the set of basis functions needs to provide sufficient detail at the bright sources
while ensuring a smooth behavior over the entire field of view, including the higher order terms
The mathematical derivation for both improvements is presented here.
The implementation is still need to be done.

\section{ACKNOWLEDGEMENTS}
This publication is part of the project LOFAR Enhanced Network for Sharp Surveys (LENSS) 
with file number 175.2023.005 of the research programme Research Infrastructure (RI): 
national consortia 2023/2024, which is (partly) financed by the Dutch Research Council (NWO)
 under the grant ID https://doi.org/10.61686/JRLCK64714
\bibliography{report} 
\bibliographystyle{spiebib} 

\end{document}